\documentclass[preprint,prb,aps,amssymb,amsmath,superscriptaddress]{revtex4}

\usepackage{color,graphicx,ulem}
\usepackage{bm}

\newcommand{\pd}[2]{\ensuremath{ \frac{\partial #1} {\partial #2} } }

\renewcommand{\eqref}[1]{Eq.\ (\ref{#1})}
\newcommand{\eqsref}[2]{Eqs.\ (\ref{#1}) and (\ref{#2})}
\newcommand{\eqsdash}[2]{Eqs.\ (\ref{#1})--(\ref{#2})}

\newcommand{\secref}[1]{Sec.\ \ref{#1}}
\newcommand{\apref}[1]{Appendix \ref{#1}}

\newcommand{\dg}{\cdot\nabla}      

\newcommand{\phb}{{\bar{\varphi}}}
\newcommand{\thb}{{\bar{\theta}}}
\newcommand{\pht}{{\varphi}}
\newcommand{\tht}{\theta}
\newcommand{\phh}{{\hat{\varphi}}}
\newcommand{\thh}{{\hat{\theta}}}

\newcommand{\tcite}[1]{Ref.\ \onlinecite{#1}}

\begin{document}

\title{Resonant Excitation of Shear Alfv\'en Perturbations by Trapped Energetic Ions in a Tokamak}
\author{}
\affiliation{
	JET-EFDA, Culham Science Centre, Abingdon OX14 3DB, UK
}
\author{I.\ G.\ Abel}
\email{Ian.Abel@jet.uk}
\affiliation{
	Rudolf Peierls Centre for Theoretical Physics, University of Oxford, Oxford OX1 3NP, UK
}
\affiliation{
	Euratom/UKAEA Fusion Association, Culham Science Centre, Abingdon OX14 3DB, UK
}
\author{B.\ N.\ Breizman}
\affiliation{
	Institute for Fusion Studies, The University of Texas, Austin, Texas 78712
}
\author{S.\ E.\ Sharapov}
\affiliation{
	Euratom/UKAEA Fusion Association, Culham Science Centre, Abingdon OX14 3DB, UK
}
\author{JET-EFDA contributors}
\thanks{See Appendix F of F.~Romanelli et. al. in Proceedings of the 22\textsuperscript{nd} IAEA Fusion Energy Conference, Geneva, 2008}
\date{\today}
\begin{abstract}
A new analytic expression is derived for the resonant drive of high $n$ Alfv\'enic modes by particles accelerated to high energy by Ion Cyclotron Resonance Heating. This derivation includes finite orbit effects, and the formalism is completely non-perturbative. The high-$n$ limit is used to calculate the complex particle response integrals along the orbits explicitly. This new theory is applied to downward sweeping Alfv\'en Cascade quasimodes completing the theory of these modes, and making testable predictions. These predictions are found to be consistent with experiments carried out on the Joint European Torus [P. H. Rebut and B. E. Keen, Fusion Technol. 11, 13 (1987)].
\end{abstract}
\maketitle
\section{Introduction}

The behaviour of highly energetic ions in tokamak plasmas is one of the most important issues in the physics of burning plasmas\cite{iterphysics5,progiterphysics5}, and thus of great interest for both theoretical and experimental study. Alfv\'enic instabilities which can tap the free energy in the energetic ion distribution~\cite{mikhailovskii1975tdi} are often excited in present-day tokamak experiments. This is due to the resonant interaction between shear Alfv\'en waves and ions either accelerated by ion-cyclotron resonance heating (ICRH)~\cite{wong1999rae} or produced by neutral beam injection (NBI)~\cite{wong1999rae}.
It is anticipated that similar Alfv\'enic instabilities will be excited in burning plasmas such as ITER by the fusion-produced alpha particles~\cite{iterphysics5,progiterphysics5}. 
The destabilization of Alfv\'en Eigenmodes (AEs) typically results in spatial redistribution (potentially global) of the fast particles~\cite{sigmar:1506}, which may lead to quenching the plasma burn and damaging the first wall. More recently AEs have also been used for diagnostic purposes~\cite{sharapov2001mhds}, divulging information about the plasma interior inaccessible to other diagnostics.

In order to determine the potential effects of AEs and what information they can reveal we require a complete model for both the modes and their interaction with the energetic ions.
In the analytic study of energetic particle driven instabilities the focus has often been on the excitation of bulk plasma eigenmodes\cite{rosenbluthrutherford,tsang:1508,breizman1995epd,fulop1996fow}. 
There is also a lot of interest in instabilities associated with non-perturbative energetic particle modes~\cite{zonca:4600,chen1984eik}. 
We continue this progress here by developing a new analytic model for the resonant drive of shear Alfv\'en perturbations due to ICRH-accelerated ions, suitable for analyzing both classes of instabilities.

The distribution of ions created by ICRH-acceleration is a distribution of particles with predominantly perpendicular energy and sharply peaked in pitch angle~\cite{stix1975fwh}. The spatial distribution of these particles is such that all the tips of the banana orbits lie on the surface where the local ion cyclotron frequency resonates with the applied RF wave. In a typical tokamak this is an approximately vertical surface, $R\approx R_{RF}$, and we classify the heating by where $R_{RF}$ lies compared to the magnetic axis $R_0$. If $R_{RF} < R_0 $ we have high-field-side heating, $R_{RF} > R_0$ gives low-field-side heating and $R_{RF} = R_0$ is known as on-axis heating. In a typical discharge with ICRH the flux-surface-averaged density of fast particles can be hollow for low-field-side heating, and is usually centrally peaked when there is high-field-side or on-axis heating. Given the energetic particle distribution we can calculate the fast particle effect on any given mode.

Our considerations will be limited to the case of Alfv\'enically-polarized fluctuations with high toroidal mode number ($n \gg 1$). In \secref{acq} we use this approximation to amalgamate previous results and produce a linearized MHD description of the problem including a kinetic contribution from the energetic particles through their perturbed anisotropic pressure.

\begin{figure}
\centering
\includegraphics[width=80mm]{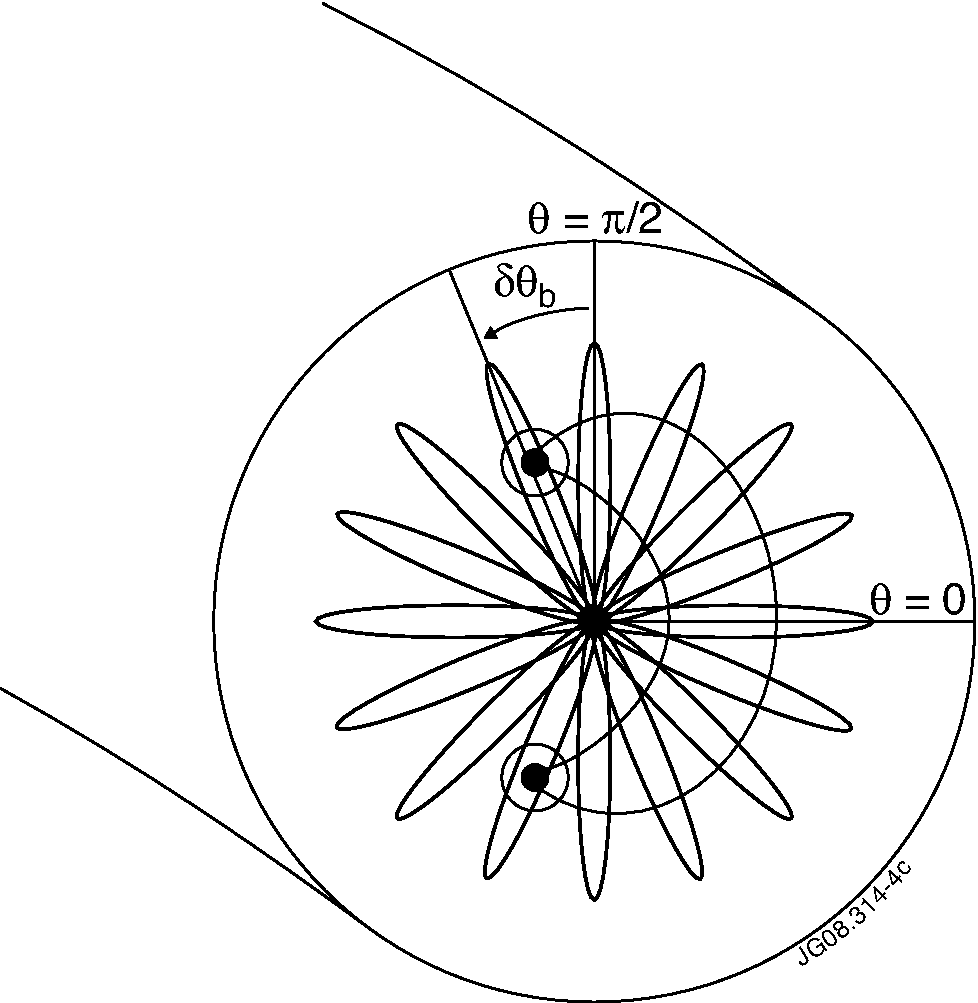}
\caption{Schematic mode structure, depicted by one equipotential surface, superposed with the poloidal cross section of an energetic particle banana orbit}
\label{modestruct}
\end{figure}

In \secref{canonical} , buliding on previous work \cite{berk1995saw,wong:2781,fulop1996fow} , we cast the wave-particle resonance in Hamiltonian form and explicitly determine the transformation to a set of action-angle coordinates for the
complex bounce-precessional energetic particle orbits. 
In \secref{fpp} we use these coordinates to express the linearized drift kinetic equation for the fast particles in a simple form. We then solve this equation for the fast particle response and use it to calculate the contribution that the fast particle resonance provides to the Alfv\'en wave equations developed in \secref{acq}. The result is expressed in terms of complex integrals over the unperturbed orbits, to make further progress we focus on the case of radially extended perturbations, for which
\begin{equation}
\label{ddrsmall}
 \pd{}{r} \ll \frac{m}{r}
\end{equation}
(where $r$ is the minor radius of the plasma and $m$ is the poloidal mode number). This approximation is valid for high-$n$ Alfv\'enic perturbations if the magnetic shear is sufficiently low over a substantial fraction of the minor radius. 
The radial structure of the mode is known to be sensitive to the radial variation of the $q$-profile.  The Alfv\'en frequency variation is itself controlled by the $q$-profile variation
which is amplified  by the large mode number. Thus if the low shear region extends for multiple poloidal wavelengths the relation in \eqref{ddrsmall} should hold.
The envisioned mode structure is shown schematically in Figure.~\ref{modestruct}.

This assumption and the thin orbit approximation imply the mode is much broader than the orbit width, which alllows us to average over a radial distance wider than an orbit width yet smaller than the mode width. The sheared magnetic field configuration leads to the result that the energetic particles experience different phases of the mode at different radii along their orbit; as these differences are multiplied by the large mode number $n$ they lead to highly oscillatory contributions to the integrals along the energetic particle orbits. The averaging procedure described allows us to calculate the integrals along orbits in closed form by only considering the non-oscillatory contributions in the vicinity of the turning points. This enables us to formulate a local expression for the resonant particle response, which can then be used in a variety of problems associated with ICRH-driven Alfv\'enic modes.

Many modes satisfy the condition in \eqref{ddrsmall}, including Low-Shear Toroidal Alfv\'en Eigenmodes~\cite{fu:1029,berk:3401,candy1996mlt}, which appear as ``tornado'' modes in experiments~\cite{kramer2004oot,sandquist:122506}. \eqref{ddrsmall} is also applicable to the Alfv\'en Eigenmodes observed in reversed shear discharges on JET~\cite{sharapov2001mhds,sharapov2002awc}, JT-60U~\cite{kimura1998aee} and Alcator C-Mod~\cite{snipes:056102}. In order to demonstrate the utility of our technique for performing resonant drive calculations we apply it in \secref{rsaedrive} to the problem of the Alfv\'en Cascade (AC) modes in scenarios with weakly-reversed shear. An explanation for the downward sweeping mode observed in such scenarios as a weakly-damped propagating quasimode has been proposed~\cite{borisvarenna}, which we complete by calculating the radiative damping and including the resonant contribution from the energetic ions. 
The damping and the drive can be evaluated for both downward and upward sweeping modes. We culminate in \secref{exp} by comparing our theoretical results with experimental observations from the JET experiment~\cite{rebut1987jet}.
\section{Alfv\'enic Perturbations in a Tokamak}
\label{acq}

Motivated by fusion applications, we will make the following assumptions from the outset. Firstly, in all cases $\beta$ (the ratio of thermal pressure to magnetic pressure) will be considered to be vanishingly small ($\beta \ll 1/n$), secondly we look at modes elongated along the field line, i.e. having parallel wavenumbers $k_\parallel = (qR)^{-1} \left( m - nq \right)$ much smaller than perpendicular wavenumbers $k_\perp \approx m / r$, finally we will make the approximation of small inverse aspect ratio ($\epsilon = r / R \ll 1$) and shifted circular flux surfaces where necessary. 

Previous works on AEs have computed the contribution to the linear MHD equations governing Alfv\'enic perturbations due to an anisotropic pressure~\cite{berk2001tia,berk1992cdl}, due to pressure gradients in the thermal plasma~\cite{fu2003epg} and due to coupling of the shear Alfv\'en perturbations to acoustic and compressional Alfv\'en perturbations~\cite{breizman2005ppe}. Assuming a background plasma with magnetic field $\bm{B}$, pressure $P$ and density $\rho$ and that the toroidal mode number $n$ of the mode is large we can combine these results to give the following wave equation:

\begin{equation}
\begin{split}
\label{modeequation}
\frac{1}{v_A^2} \nabla^2_\perp \pd{^2 \Phi}{t^2} &- \bm{B} \dg \left(\frac{1}{B^2} \nabla_\perp^2 \bm{B}\dg \Phi \right) \\ 
&= \frac{Z e}{c} \int d^3 \bm{v} \bm{v}_D \dg \delta f + \frac{8\pi P}{B^2}\bm{b}\times \bm{\kappa}\dg \left[ \left( \bm{b}\times\nabla \ln P - \gamma_H \nabla\times \bm{b}\right)\dg \Phi \right] ,
\end{split}
\end{equation}
where $v_A = B / \sqrt{4\pi \rho}$ is the local Alfv\'en speed, $\bm{b} = \bm{B}/B$ the magnetic field direction, $P$ the thermal plasma pressure, $\bm{\kappa} = \bm{b}\dg\bm{b}$ the curvature, $\bm{v}_D = \frac{1}{M\Omega} \left( \mu \nabla B + M v_\parallel^2 \bm{b}\times\bm{b}\dg\bm{b}\right)$ the fast particle drift velocity, $\mu = M v_\perp^2 / 2B$ the magnetic moment and $\gamma_H$ the ratio of specific heats of the background plasma. The fast particles' perturbed distribution function is $\delta f$ and their mass is $M$, charge $Ze$ and cyclotron frequency $\Omega = Z e B / M c$.
We have followed \tcite{breizman2005ppe} and expressed the electromagnetic perturbation as
\begin{align}
\label{empert}
\delta \bm{B} &= \nabla\times \left(\left(\bm{b} \times\nabla \Phi\right)\times \bm{b}\right),\\
		\label{empert2}
\delta\bm{E}  &= -\frac{1}{c}\nabla_\perp \pd{\Phi}{t}.
\end{align}
This formulation of the problem relies on $\beta \ll 1$ and $k_\parallel \ll k_\perp$ to eliminate the coupling to the compressional Alfv\'en wave (fast magnetosonic mode), and on $2 q^2 \gg 1$ to neglect the coupling to acoustic perturbations (this is equivalent to assuming that the Geodesic Acoustic Mode (GAM) is supersonic). This decoupling is carried out in detail in \tcite{breizman2005ppe}.

To	compute the fast particle resonant contribution to \eqref{modeequation}, we express the integral over velocity space as a curvature coupling between the Alfv\'enic perturbation and the perturbed anisotropic fast particle pressure (again using the fact that $\beta$ is small to simplify the drift velocity),
\begin{align}
\label{vdpperp}
\frac{Ze}{c} \int d^3 \bm{v} \bm{v}_D \dg \delta f = \frac{4\pi}{B^2} \bm{b}\times\bm{\kappa}\dg \delta P_\perp,
\end{align}
where we have neglected the small parallel pressure in comparison to the perpendicular pressure. 
We now merely have to complement \eqref{modeequation} by expressing the perturbed fast particle pressure $\delta P_\perp$ in terms of $\Phi$. This has already been done in \tcite{berk2001tia} for the non-resonant component of $\delta P_\perp$ , thus we only need consider the resonant component here.

\section{Hamiltonian Form of the Unperturbed Orbits}
\label{canonical}
In this section we reduce the dynamics of the particles to a completely integrable Hamiltonian form, keeping finite orbit width effects but neglecting finite Larmor radius (FLR) effects.
We start from the Hamiltonian form of the Littlejohn Lagrangian~\cite{littlejohn1983} as presented in \tcite{berk1995saw},
\begin{align}
L &= P_\theta \dot{\theta} + P_\varphi \dot{\varphi} + P_\zeta \dot{\zeta} - H( P_\varphi, P_\theta, \theta, P_\zeta ), \\
P_\zeta &= \frac{Mc}{Ze} \mu,\\
\label{ptheta}
P_\theta &= \frac{Ze}{c} A_\theta + M v_\parallel \frac{B_\theta}{B}\\
\label{pphi}
P_\varphi &= \frac{Ze}{c} A_\varphi + M v_\parallel \frac{B_\varphi}{B}, \\
H &= \frac{Ze}{cM} P_\zeta B + \frac{1}{2M} \left( P_\varphi - \frac{Ze}{c} A_\varphi \right)^2 \left( \frac{B}{B_\varphi} \right)^2.
\end{align}
Here subscripts denote the covariant component of a vector quantity. In addition $\zeta$ is the gyroangle about the equilibrium field and we have picked orthogonal coordinates $r$, $\tht$, $\pht$ corresponding to flux surface label, poloidal angle and toroidal angle respectively. The definition of these coordinates is chosen such that $B_r$ vanishes (equivalently $r$ is orthogonal to $\tht$). We have also picked a gauge such that the unperturbed $A_r$ vanishes. This is critical to being able to write the $\bm{v}\cdot\bm{\dot{x}}$ term in the Lagrangian in Hamiltonian form~\cite{berk1995saw,meiss1990ccg}. 
It is understood that all quantities should now be viewed as functions of $P_\zeta, P_\tht, P_\pht$ and $\zeta, \tht, \pht$ as one can define $r$ and $v_\parallel$ implicitly from \eqsref{pphi}{ptheta}. 

In order to express the motion in action angle form, we follow \tcite{berk1995saw} and introduce a generating function $g$ for the transformation to new canonical coordinates $\phb$, $\thb$, $\bar{\zeta}$ and momenta $P_\phb$, $P_\thb$, $P_{\bar{\zeta}}$. We now consider a function $g$ that leaves $P_\pht$ and $P_\zeta$ unchanged i.e.
\begin{equation}
g = P_{\phb} \pht + P_{\bar{\zeta}} \zeta + g^*(P_\phb ; P_\thb ; P_{\bar{\zeta}} ; \theta ),
\end{equation}
which gives the following equations for $\phb$ and $\thb$:
\begin{align}
\label{phbd}
\phb &= \pht + \pd{g^*}{P_\phb}, \\
\label{thbd}
\thb &= \pd{g^*}{P_\thb}.
\end{align}
The function $g^*$ is determined from
\begin{equation}
\label{defg}
P_\tht = \pd{g^*}{\theta}.
\end{equation}
In order to solve this we need an explicit form for $P_\tht$ as a function of $\tht$ and conserved quantities $P_\phb$, $P_\thb$, $P_{\bar{\zeta}}$ or equivalently $P_\pht$, $P_\zeta$ and $H$. This nessecitates picking an explicit form for both $\bm{A}$ and $\bm{B}$ subject to the orthogonality constraint mentioned above.

In a large aspect ratio tokamak we can do this by expanding all quantities in the inverse aspect ratio to find~\cite{meiss1990ccg},
\begin{align}
\label{aphi}
A_\varphi &= -B_0 \int\limits_0^r \frac{r d r}{q(r)},\\
A_\theta  &= \frac{1}{2} B_0 r^2 - B_0 \alpha(r) \cos \theta  ,\\
B_\varphi &= B_0 R_0,\\
B_\theta  &= B_0 \frac{r^2}{R_0 q(r)} \left( 1 - \frac{\alpha'(r)}{r} \cos \theta \right),
\end{align}
with the function $\alpha$ defined by
\begin{align}
\alpha(r) = \int\limits_0^r \left(\frac{{r}^2} {R_0} + r \Delta' + r \int\limits_0^{r} \frac{\Delta'(s)}{s} ds\right) dr,
\end{align}
with primes denoting radial derivatives, $\Delta(r)$ the Shafranov shift , $s$ a dummy variable, $B_0$ the magnetic field strength on axis and $R_0$ the major radius of the magnetic axis.
We also note that there is a simple transformation from orthogonal coordinates to straight-field-line coordinates $r,\thh,\phh$ which are convenient for describing Alfv\'enic modes (see Ref.~\onlinecite{meiss1990ccg}),
\begin{align}
\thh &= \tht - \frac{\alpha'(r)}{r} \sin\theta,\\
\phh &= \pht.
\end{align}
In what follows we replace $P_\varphi$ by a new radial coordinate $\bar{r}$ such that
\begin{align}
\label{pphir}
P_\varphi &\equiv -\frac{Ze}{c} B_0 \int\limits_0^{\bar r} \frac {r \, dr}{q(r)} .
\end{align}
We thus have $P_\theta$ given by \eqref{ptheta} as a function of $r$, $\theta$ and $v_\parallel$ and must use \eqsref{pphir}{pphi} to eliminate $r$ in favour of $\bar{r}$, $P_\zeta$ and $\theta$. 
We start by neglecting the FLR correction in \eqref{ptheta}, finding that $P_\tht$ is now a function only of the spatial coordinates,
\begin{equation}
\label{ptheta2}
P_\theta \approx \frac{Ze}{c} A_\theta.
\end{equation}
We now make use of the fact that we are dealing with ICRH-accelerated particles, which are poloidally trapped between bounce points where $v_\parallel$ changes sign. The majority of these particles perform banana orbits where they deviate only slightly from the flux surface of the bounce points, thus we write,
\begin{align}
r &= \bar{r} + \delta,
\end{align}
where $\delta$ is the deviation from the flux surface and we identify $\bar{r}$ as the radius of the bounce points.
We will now assume $r,\bar{r} \gg \delta$. This means that in most places we can replace $r$ with $\bar{r}$, as effects of the order of $\delta / r$ are the smallest effects we wish to keep (finite orbit width, but not FLR effects). However in doing so we limit ourselves to orbits that do not pass close to the magnetic axis as $r \sim \delta$ there, and such orbits are no longer banana orbits but so-called potato orbits. The thin orbit approximation allows us to find $\delta$ explicitly and eliminate $r$ in favour of $\bar{r}$ and $\delta$ as required. In the potato orbit regime a different approximation must be used to eliminate $r$. As the AC modes under consideration here are well away from the magnetic axis we make the thin-orbit approximation in all further analysis. 

We thus obtain,
\begin{align}
\label{rrbar}
  r = \bar{r} + \frac{R_0 q(\bar{r})}{\bar{r}} \frac{v_\parallel}{\Omega}.
\end{align}
We also introduce the bounce angle $\theta_b$, being $\theta$ at the $v_\parallel =0$ point, which we will use instead of $P_\thb$.
Eliminating $r$ we find,
\begin{align}
P_\tht &=  \frac{Ze}{c} \left(  \frac{1}{2} B_0 \bar{r}^2 - B_0 \alpha(\bar{r}) \cos \theta \right) + q(\bar{r}) \sqrt{{R_0}{\bar{r}}}\sqrt{\frac{2Ze}{c} P_\zeta B_0 }\sqrt{\cos \theta - \cos \theta_b} .
\end{align}
We can now solve \eqref{defg} for $g^*$ by substituting the following ansatz,
\begin{equation}
\label{gsol}
g^* = \frac{Z e B_0}{c} \left( \frac{1}{2} \bar{r}^2 \theta - \alpha(\bar{r}) \sin\tht \right) + q(\bar{r}) \sqrt{R_0\bar{r}}\sqrt{\frac{2Ze}{c} P_\zeta B_0 } G,
\end{equation}
where $G$ now satisfies 
\begin{equation}
\label{badger}
\left( \frac{dG}{d\tht} \right)^2 = \cos \tht - \cos \theta_b .
\end{equation}
Using \eqref{gsol} in \eqsref{phbd}{thbd} results in:
\begin{align}
\label{phbdef}
\phb = \varphi - q \tht + \frac{q \alpha'}{\bar r} \sin \tht - G(\theta) \chi_1(\bar{r}) + \left(\frac{\omega_\phb}{\omega_{\thb}} - \chi_2(\bar{r})\right) \thb, \\
\label{thbdef}
\thb = \omega_\thb { \bar{r} q } \sqrt{\frac{2M}{\Omega P_\zeta }} \left(\frac{ R_0 }{\bar{r}} \right)^{3/2} \frac{1}{\sin \tht_{b}}  \pd{G}{\tht_b},
\end{align}
where the functions $\chi_1(\bar r)$ and $\chi_2(\bar r)$ are defined as follows,
\begin{align}
\chi_1 = \frac{q(\bar r)}{\bar r} \pd{}{\bar r} \sqrt{  \frac{2P_\zeta}{M\Omega} {\bar r} R_0 q^2({\bar r}) } \\
\chi_2 = \frac{ P_\zeta \cos \theta_b }{M\omega_\thb  R_0 {\bar r} } q(\bar{r})
\end{align}
The general solution of \eqref{badger} is,
\begin{equation}
\label{Gdef}
G = \pm 2^{3/2} \sin \left(\frac {\theta_b} {2} \right) \mbox{E}\left( \frac{\theta}{2} \right|\left. \csc^2 \left(\frac{\theta_b}{2} \right) \right) + A,
\end{equation}
with $E(z | m)$ the elliptic integral of the second kind with modulus $m$~\cite{abramowitz1965hmf}, and $A$ an arbitrary constant. 
Choosing our orbit to start from $\tht = -\theta_b,\thb=-\pi$, pass through $\tht = \tht_b,\thb=0$, and end at $\tht=-\tht_b,\thb=\pi$ determines the constants and the ambiguous signs, giving:
\begin{align}
\label{Gb1}
G &= &2^{3/2} \sin \left(\frac {\theta_b} {2} \right) \left\{\mbox{E}\left( \frac{\theta}{2} \right|\left. \csc^2 \left(\frac{\theta_b}{2} \right) \right) - \mbox{E}\left( \frac{\theta_b}{2} \right|\left. \csc^2 \left(\frac{\theta_b}{2} \right) \right)\right\}&\mbox{ when } \thb \in (-\pi,0),\\
\label{Gb2}
G &= -&2^{3/2} \sin \left(\frac {\theta_b} {2} \right) \left\{\mbox{E}\left( \frac{\theta}{2} \right|\left. \csc^2 \left(\frac{\theta_b}{2} \right) \right) - \mbox{E}\left( \frac{\theta_b}{2} \right|\left. \csc^2 \left(\frac{\theta_b}{2} \right) \right)\right\}&\mbox{ when } \thb \in (0,\pi),
\end{align}
and
\begin{align}
\label{dthetadtheta}
\frac{d \thb}{d \theta} = ~~\thinspace\omega_\thb \frac{R_0 q}{|v_\parallel|}&\mbox{ when } \thb \in (-\pi,0),\\
		\label{dthetadtheta2}
\frac{d \thb}{d \theta} = -\omega_\thb \frac{R_0 q}{|v_\parallel|}&\mbox{ when } \thb \in (0,\pi),
\end{align}
with
\begin{equation}
\label{vpar}
v_\parallel = \sqrt{\frac{ 2 B_0 \bar{r}}{R_0 M^2}\frac{Ze}{c} P_\zeta} \frac{d G}{d \tht}.
\end{equation}
This completes the description of the transformation to the coordinates $\thb,\phb$ from the orthogonal coordinates $\tht,\pht$.

To complete the analysis of the orbits, we must determine the frequencies with which the particle performs its periodic motion in $\thb$, $\phb$.
These frequencies however are merely the particle bounce frequency and toroidal precession frequency which in the large aspect ration limit are given by,~\cite{white2006ttc}
\begin{align}
\label{bouncefreq}
\omega_\thb &= \frac{\pi}{2 R_0 q(\bar{r}) K(\kappa)}\sqrt{\frac{\bar{r}\mu B_0}{M R_0}} \\
\label{precessfreq}
\omega_\phb &= \frac{c}{Ze} \frac{\mu q(\bar{r})}{R_0\bar{r}} \left\{{4s(\bar{r})}\frac{E(\kappa) + \left(\kappa^2 - 1\right) K(\kappa)}{K(\kappa)} + \frac{2E(\kappa) - K(\kappa)}{K(\kappa)}\right\}
\end{align}
with $\kappa = \sin (\tht_b / 2)$ and $K$ and $E$ the complete elliptic integrals of the first and second kinds respectively. We also write the magnetic shear as
\begin{align}
\label{shear}
s(r) = \frac{r}{q(r)} \frac{d q}{d r}.
\end{align}
\section{Calculation of the Fast Particle Resonant Response}
\label{fpp}
Having found action angle coordinates in which the unperturbed motion of the energetic particles is completely integrable, we now turn to the problem of calculating the energetic particle response to shear Alfv\'en perturbations. Similar derivations for perturbative modes and deeply trapped particles have been done before in \tcite{fulop1996fow} and \tcite{wong:2781}.
For the case we will consider in \secref{rsaedrive} we expect the resonant instability drive to be small compared to the wave frequency and thus for simplicity we will only consider the leading order contribution to the resonant fast particle distribution function in the current section.
We start by writing the collisionless kinetic equation for the energetic particle distribution function as
\begin{equation}
\frac{d f}{dt} = 0 ,
\end{equation}
with the time derivative being taken along the particle orbit. This can then be linearized by expanding the distribution function as $f = F + \delta f$ where $\delta f $ is the perturbation due to the mode and $F$ is an equilibrium distribution function that only depends on the constants of motion. The resulting equation is
\begin{equation}
\label{ldke}
\frac{d}{dt} \delta f = - \frac{d}{dt} F,
\end{equation}
with the derivative on the left evaluated along the unperturbed orbits, and the one on the right along the perturbed orbits. We now write the left hand side of \eqref{ldke} using the coordinates found in \secref{canonical} and average over the fast motion in $\zeta$ to obtain
\begin{equation}
\label{lhsdke}
\frac{d}{dt} \delta f = \left(\pd{}{t} + \omega_\phb\pd{}{\phb} + \omega_\thb \pd{}{\thb} \right) \delta f. 
\end{equation}
If we pick $F=F(U,\mu,P_\tht)$, where $U=\frac{1}{2}mv^2$ is the particle energy, then the right hand side of \eqref{ldke} can be evaluated by conventional means~\cite{berk1995saw,simakov2005dke,hazeltine1973rdd} for the electromagnetic perturbation given by \eqsref{empert}{empert2}, which results in
\begin{equation}
\label{rhsdke}
\frac{d}{dt} F = \frac{Ze}{c} \pd{}{\pht} \left(\pd{\Phi}{t}\right) \pd{F}{P_\pht} - \frac{Ze}{c} \bm{v}_D \dg \pd{\Phi}{t} \pd{F}{U},
\end{equation}
where derivatives of $F$ are taken holding $\mu$ and either $U$ or $P_\tht$ constant. We note that the second term on the right hand side of \eqref{rhsdke} is smaller than the first term by a factor of $r/R$ and thus we neglect it and focus on radial gradients of the energetic particle distribution as the main instability driving mechanism.
Combining \eqsref{lhsdke}{rhsdke} results in
\begin{equation}
\label{finaldke}
\left(\pd{}{t} + \omega_\phb\pd{}{\phb} + \omega_\thb \pd{}{\thb} \right) \delta f = -\frac{Ze}{c} n \omega \sum_m {\Phi}_m e^{in\phh-im\thh - i \omega t} \pd{F}{P_\pht},
\end{equation}
where we have made use of a Fourier expansion of $\Phi$ in time and in straight-field-line coordinates ($\phh$,$\thh$) as follows:
\begin{equation}
\label{fourierphi}
\Phi = \sum_m {\Phi}_m (r) e^{i n \phh - i m \thh-i\omega t} .
\end{equation}
Axisymmetry implies (due to \eqref{phbdef}) that Fourier harmonics in $\phb$ are decoupled, and thus one $\phh$ harmonic only gives rise to one $\phb$ harmonic.
Expanding $\delta f$ in a series of $\thb$ harmonics and a single $\phb$ and time harmonic then gives an immediate solution of \eqref{finaldke} :
\begin{equation}
\label{deltafsoln}
\delta f  = -i n\omega \sum\limits_{l,m} \frac{Ze}{c} \frac{{\Phi}_m(\bar{r}) K(l,m) e^{in\phb-il\thb - i \omega t}}{\omega - n \omega_\phb + l\omega_\thb} \pd{F}{P_\varphi}. 
\end{equation}
To transform from $r, \thh, \phh$ straight-field line coordinates to $\bar{r}, \thb, \phb$ coordinates, we have made use of the slow radial variation of $\Phi_m$ to 
replace $\Phi_m(r)$ by $\Phi(\bar{r})$ and defined the internal integral $K$, which transforms straight-field-line Fourier harmonics to action-angle Fourier harmonics, by
\begin{equation}
\label{kdef}
K(l,m) = \frac{1}{4\pi^2} \int \limits_0^{2\pi} d\phb \int\limits_{-\pi}^{\pi} d\thb e^{in\phh - im\thh}e^{-in\phb+il\thb}.
\end{equation}
In general, the evaluation of $K$ involves a complex integral over the entire orbit. However, in the high-$n$ approximation, with $k_\parallel \ll k_\perp$,
we can evaluate $K$ asymptotically, which we do in \apref{apintegrals}.

 We now use $\delta f$ as given by \eqref{deltafsoln} to compute $\delta P$ for inclusion in the mode equation (\eqref{modeequation}), via \eqref{vdpperp}.
In order to integrate $\delta f$, we write integrals over all velocities as integrals over $U, \mu$ and $\zeta$. However it is important to note that
whilst in the definition of the pressure the velocity integration is done at constant $r$, $\delta f$ is a function of $\bar{r}$. We make use of the thin orbit apprixmation and
slow radial variation of all quantities to replace all instances of $\bar{r}$ with $r$ in the integrand, and perform all integrals at constant $r$.
Thus
\begin{equation}
\int d^3 \bm{v} = \sum_{\sigma} \int dU d\mu d\zeta \frac{B}{M^2|v_\parallel|} \approx \sum_\sigma \int dU d\mu d\zeta \frac{ B_0 }{M R_0 q(\bar{r}) \omega_\thb} \left|\frac{d \thb}{d \tht}\right|,
\end{equation}
where \eqref{dthetadtheta} has been used to replace $|v_\parallel|$, and $\sigma \equiv v_\parallel / |v_\parallel|$.
This simplifies the expression for $\delta P_\perp$ to
	  \begin{equation}
	  \delta P_\perp = 2 \pi\sum\limits_\sigma \int dU d\mu \frac{ B_0^2 }{M^2 R_0 q(\bar{r}) \omega_\thb} \left|\frac{d \thb}{d \tht} \right|\mu \delta f.
		\end{equation}
		Subsituting $\delta f$ from \eqref{deltafsoln} we then have
	\begin{equation}
	  \delta P_\perp = -2\pi i n \omega \Omega \frac{B_0}{M R_0 q } \sum_{l,m,\sigma} {\Phi}_m \int dU d\mu  \left|\frac{d \thb}{d \tht}\right| \frac{\mu}{\omega_\thb} \frac{K(l,m) e^{in\phb-il\thb-i\omega t}}{\omega - n\omega_\phb + l\omega_\thb} \pd{F}{P_\varphi}.
		\end{equation}	
		We now expand $\delta P_\perp$ as a Fourier series in $\phh$ and $\thh$, and interchange integration over $\phh$ and $\thh$ with integration over $U$ and $\mu$ to find
		\begin{align}
		\delta P_\perp &= \sum_{j} {\delta P}_{\perp j} (r) e^{in\phh - ij\thh - i\omega t}\\
		\label{dp}
	{\delta P}_{\perp j} &= -2\pi i n \omega \Omega \frac{ B_0 }{M R_0 q }  \sum_{l,m} {\Phi}_m\int dU d\mu  \frac{\mu}{\omega_\thb} \frac{ K(l,m) K^{ext}(l,j) }{\omega - n\omega_\phb + l\omega_\thb} \pd{F}{P_\varphi}.
	\end{align}
	We have defined the external integral $K^{ext}$, which is similar to $K$ except involved with the inverse transformation back to straight-field-line harmonics, by
	\begin{equation}
	\label{kextdef}
	K^{ext}(l,j) = \sum_\sigma \frac{1}{4\pi^2} \int \limits_0^{2\pi} d\phh \int\limits_{-\pi}^{\pi} d\thh \left|\frac{d\thb}{d\tht}\right| e^{in\phb - il\thb}e^{-in\phh+ij\thh},
	\end{equation}
	and again defer its evaluation to \apref{apintegrals}.

	We now simplify ${\delta P}_{\perp j}$ by selecting only the resonant contribution, i.e. we use the formula
	\begin{align}
	\int \frac{dx}{x} = -i \int dx \delta(x) + \mathcal{P} \int \frac{dx}{x} ,
	\end{align}
	with $\mathcal{P}\int dx$ denoting the Cauchy principal value of the integral. We define $\delta P^{R}_\perp$ to be the part of the perturbed pressure associated with the resonant delta function.
	We then use the nature of the background fast-ion distribution $F$ to perform the integration over $U$; the approximation that all the fast ions bounce at the resonant magnetic field strength (where the ICRH deposition is localized) is equivalent to factorising $F$ into
	\begin{equation}
	\label{eq57}
	F = F_0(P_\varphi,\mu,U) \delta \left( U - \mu B_{res} \right).
	\end{equation}

	We finally combine \eqsref{dp}{eq57} into
		\begin{align}
		\label{dpfinale}
	{\delta P}^R_{\perp j} &= -2\pi n \omega \Omega \frac{ B_0 }{M R_0 q } \sum_{l,m} {\Phi}_m \int d\mu \delta\left({\omega - n\omega_\phb + l\omega_\thb}\right)  \frac{\mu}{\omega_\thb} { K(l,m) K^{ext}(l,j) } \pd{F_0}{P_\pht},
	\end{align}
	with all instances of $U$ evaluated at the resonant energy $\mu B_{res}$.
		If we average this over a small radial region (larger than the orbit width but less than the radial wavelength of the mode) or equivalently neglect components of $\delta P$ that vary rapidly in radius we can use the explicit expression for $K(l)K^{ext}(j)$ from \eqref{llama} to ultimately obtain
		\begin{align}
		\label{dpuber}
		{\delta P}^R_{\perp j} &= -2\pi n \omega \Omega \frac{ B_0 }{M R_0 q } \sum_{l,m}{\Phi}_m  \int d\mu \delta\left({\omega - n\omega_\phb + l\omega_\thb}\right)  \frac{\mu}{\omega_\thb} \frac{e^{i\left(m-j\right)\theta_b} }{2 \pi^2 n^2 \chi_2^2} \pd{F_0}{P_\pht},
		\end{align}
		if $\tht_b < \frac{\pi}{2}$ or
		\begin{align}
		\label{dpuber2}
		{\delta P}^R_{\perp j} &= -2\pi n \omega \Omega \frac{ B_0 }{M R_0 q }  \sum_{l,m} {\Phi}_m \int d\mu \delta\left({\omega - n\omega_\phb + l\omega_\thb}\right)  \frac{\mu}{\omega_\thb} \frac{8}{\pi|nG^*|} \cos\left( (m - j)\tht^* \right) \pd{F_0}{P_\pht},
		\end{align}
		if $\tht_b > \frac{\pi}{2}$. All quantites in these expressions are now functions of $r$ and not $\bar{r}$ (see discussion after \eqref{kdef}). The quantities $G^*$ and $\tht^*$ are defined by
		\begin{align}
		G^* &\equiv \frac{2 P_\zeta}{M} \frac{\Omega \bar r \chi_1}{q^2 R_0^3 \omega_\thb^2} \left( - \sin \tht^* \sqrt{\cos \tht^* - \cos\tht_b}\right),\\
		\cos \tht^* &= \frac{1}{2} \cos\tht_b.
		\end{align}
	\section{Application to Alfv\'en Cascade Drive}
	\label{rsaedrive}
	In this section, we apply our analysis to the problem of Alfv\'en Cascades in shear-reversed discharges driven by ICRH-accelerated ions. 
	 We expect the resonant ions to have significantly lower energies than the non-resonant ones. We therefore use the small orbit width approach detailed above for the resonant ions and the large orbit width limit considered in \tcite{berk2001tia} for the nonresonant ions. 
	Alfv\'en Cascades can be adequately described by a single Fourier component ${\Phi}_m$ of the perturbed field (see \eqref{fourierphi}); taking the corresponding 
	component of \eqref{modeequation} results in:
	\begin{equation}
	\begin{split}
	\label{modedeq}
	\pd{}{r} &\left(\frac{\omega^2}{v_A^2} - k_\parallel^2 - \frac{\omega_G^2}{v_A^2} \right) \pd{\Phi_m}{r} - \frac{m^2}{r^2} \left(\frac{\omega^2}{v_A^2} - k_\parallel^2- \frac{\omega_G^2}{v_A^2} - \frac{\omega_\nabla^2}{v_A^2} \right) {\Phi_m} \\
			&= \frac{4\pi e m}{B_0 c r} \omega \pd{n_{0h}}{r} \Phi_m + \frac{2\pi}{i B_0 R_0} \left( \frac{m}{r} \left(\delta P^R_{\perp m+1} + \delta P^R_{\perp m-1}\right) + \pd{}{r} \left( \delta P^R_{\perp m+1} - \delta P^R_{\perp m-1} \right) \right).
	\end{split}
	\end{equation}
	We have introduced the geodesic acoustic frequency $\omega_G^2 = \frac{2}{MR^2} \left( T_e + \frac{7}{4}T_i \right)$ and the pressure gradient induced frequency shift $\omega_\nabla^2 = - \frac{2}{MR^2} r \frac{d}{dr} \left( T_e + T_i \right)$ in accordance with \tcite{borisvarenna} and \tcite{fu2003epg}.
	We can write the resonant pressure contributions using \eqref{dpfinale} as:
	\begin{align}
	\label{rsaecos}
			\frac{1}{2} &\left(\delta P^R_{\perp m+1} + \delta P^R_{\perp m-1}\right) \\\nonumber&= -2\pi n \omega \Omega \frac{ B_0 }{2 M R_0 q } {\Phi_m} \sum_{l} \int d\mu\delta(\omega-n\omega_\phb+l\omega_\thb)\frac{\mu}{\omega_\thb} K_1(l,m) \pd{F_0}{P_\varphi},\\
			\frac{1}{2} &\left(\delta P^R_{\perp m+1} - \delta P^R_{\perp m-1}\right) \\\nonumber&= -2\pi n \omega \Omega \frac{ B_0 }{2 M R_0 q } {\Phi_m} \sum_{l} \int d\mu\delta(\omega-n\omega_\phb+l\omega_\thb) \frac{\mu}{\omega_\thb} K_2(l,m) \pd{F_0}{P_\varphi},\\
			\label{k1def}
	K_1 &= \sum\limits_{s = \pm 1} K(l,m)K^{ext}(l,m+s), \\
			 \label{k2def}
	K_2 &= \sum\limits_{s = \pm 1} s K(l,m)K^{ext}(l,m+s).
	\end{align}
	With these expressions \eqref{modedeq} reduces to a second order differential equation
	for the shear Alfv\'en perturbation $\Phi_m$,
	\begin{equation}
	\label{modede}
\pd{}{r} \left( \omega^2 - \omega_A^2 \right) \pd{\Phi_m}{r} - \left(\omega^2 - \omega_A^2 - \omega_H^2 \right) \frac{m^2}{r^2} \Phi_m = iD_1 \Phi_m + D_2 \pd{\Phi_m}{r},
\end{equation}
where we have abbreviated the fast particle terms by defining $D_1$ and $D_2$ as follows:
\begin{align}
\frac{2\pi v_A^2}{i B_0 R_0} \frac{m}{r} \left(\delta P^R_{\perp m+1} + \delta P^R_{\perp m-1}\right) &= i D_1 \Phi_m, \\
\frac{2\pi v_A^2}{i B_0 R_0} \pd{}{r} \left( \delta P^R_{\perp m+1} - \delta P^R_{\perp m-1} \right) &= D_2 \pd{}{r} \Phi_m,
\end{align}
and defined frequencies of the Alfv\'en continuum ($\omega_A$) and the offset ($\omega_H$) from the continuum as follows:
\begin{align}
\omega_A^2 &= k_\parallel^2 v_A^2 + \omega_G^2, \\
				  \label{omHdef}
\omega_H^2 &= -\frac{\omega}{m} \Omega \frac{r}{n_i} \pd{n_{0h}}{r} + \omega_\nabla^2.
\end{align}
Before engaging in a detailed analysis of \eqref{modede}, we can immediatly determine several general features of the modes. Making a local approximation, and neglecting the small radial derivatives we have
\begin{equation}
\label{eq81}
\left({\omega}^2 - \omega_A^2 - \omega_H^2 \right)  = -A i \pd{F}{\bar{r}} K_1,
\end{equation}
where $A$ is a positive constant, and $K_1$ is defined by \eqref{K1def}. This leads to the conclusion that the fast particle contribution is destabilizing if:
\begin{itemize}
\item{$\theta_b > \frac{\pi}{2}$ and $\pd{F}{r} < 0$, i.e. High field side ICRH and a peaked radial profile,}
\item{$\theta_b < \frac{\pi}{2}$ and $\pd{F}{r} > 0$, i.e. Low field side ICRH and a hollow radial profile.}
\end{itemize}
As these are the expected profiles of ICRH ions, we expect the mode to be observed independent of heating location. We also note that these results are for $n > 0$. Looking at the exact expression in \eqref{rsaecos} we see that the destabilizing term is proportional to $n$.
Therefore we do not expect any modes with negative $n$ for normal profiles of ICRH-accelerated ions.
Returning to \eqref{modede}, we proceed as in \tcite{borisvarenna} and use our assumption of small radial gradients (\eqref{ddrsmall}) to determine that the second term on the left hand side of \eqref{modede} is larger than all the other terms. This results in
\begin{align}
\omega^2 \approx \omega_0^2 = \omega_A^2 + \omega_H^2,
\end{align}
i.e. the real frequency of the mode must be close to the local Alfv\'en frequency with the offset determined by the nonresonant fast particle density gradient and plasma pressure gradient. This means that we can neglect the small spatial dependence in the radial derivative term, 
\begin{equation}
\label{modeeq}
\pd{^2}{r^2} \Phi_m - \frac{m^2}{r^2} \frac{ \omega^2 - \omega_A^2(r) - \omega_H^2(r)}{\omega_H^2(r_0)} \Phi_m = \frac{1}{\omega_H^2(r_0)} \left(i D_1(r_0) \Phi_m + D_2(r_0) \pd{}{r}\Phi_m \right).
	\end{equation}
	We order the radial dependence of $\omega_A$ (which is predominantly from the radial variation of $q$) through
\begin{equation}
\left| \omega_A^2(r) - \omega^2_A(r_0) \right| \sim \omega_H(r_0) \frac{r^2}{m^2} \pd{^2}{r^2} .
\end{equation}
As $\omega_H \ll \omega_A$, this means that we can neglect the spatial dependence of $\omega_H^2$ everywhere. Expanding 
about $q = q_0$, the value of $q$ at the reverse shear point, where $r = r_0$ , we find (writing $\omega_0^2 = \omega_A^2(r_0) + \omega_H^2(r_0)$ to the required accuracy)
\begin{align}
\label{eqboing}
\pd{^2}{z^2} \Phi_m &= \left( \lambda  - \eta z^2 - z^4 \right) \Phi_m +  iE \Phi_m + F \pd{}{z}\Phi_m, \\
z &= \frac{r - r_0}{r_0} \left({\frac{m^2 v_A q_0'' r_0^2}{2 \omega_H q_0^2 R_0}}\right)^{1/3}, \\
\lambda &=  \frac{\omega^2- \omega_0^2}{\omega_H^2} \left({\frac{4 \omega_H^2 q_0^4 R_0^2 m^2}{v_A^2 q_0''^2 r_0^4}}\right)^{1/3}, \\
\label{etadef}
\eta &= \left(nq - m\right) \left( \frac{16 m v_A^2 }{\omega_H^2 q_0  r_0^2 q_0'' R_0^2} \right)^{1/3}, \\
E &= \frac{r_0^2}{\omega_H^2} \left( \frac{2q_0^2 \omega_H R_0}{v_A m^2 r_0^2 q''_0} \right)^{2/3} D_1, \\
F &= \frac{r_0}{\omega_H^2} \left( \frac{2q_0^2 \omega_H R}{v_A m^2 r_0^2 q''_0} \right)^{1/3} D_2.
\end{align}
We then introduce the auxiliary quantity $\Psi = e^{-Fz / 2} \Phi_m$, which reduces \eqref{eqboing} to
\begin{align}
\pd{^2}{z^2} \Psi = \left( \lambda + i E - \frac{1}{4} F^2  - \eta z^2 - z^4 \right) \Psi.
\end{align}
We also find it convenient to reintroduce explicit time dependence by writing $\Phi_m = \Phi_m(t) e^{-i\omega_0 t}$ with $\Phi_m(t)$ a slowly varying envelope. This results in
$\left(\omega^2-\omega_0^2\right) \Psi \approx 2i\omega_0 \pd{}{t} \Psi$. We can then rescale time by $t = t' \frac{2m^2\omega_0}{\omega^2_H}\left( \frac{2q_0^2 \omega_H R_0}{v_A m^2 r_0^2 q''_0} \right)^{2/3}$ to obtain the final mode equation,
\begin{align}
\label{normsch}
i\pd{}{t'} \tilde{\Psi} = \pd{^2}{z^2} \tilde{\Psi} + \left( \eta z^2 + z^4 \right) \tilde{\Psi} \\
\Psi = e^{ -  E t'  + i\frac{1}{4} F^2 t'} \tilde{\Psi}
\end{align}
which is a time-dependent Schr\"odinger equation with a quartic potential hill for large $z$ and a local hill or a well for small $z$, depending on the sign of $\eta$. The
expression for $\omega_0$ shows that a downward sweeping Alfv\'en Cascade will have $\eta > 0$ and an upward sweeping cascade will have $\eta <0$. Thus we naturally see that the upward
sweeping modes are generally real eigenmodes localized in a potential well and the downward sweeping modes are decaying quasimode solutions of \eqref{normsch} atop a potential hill.
The transformation to $\tilde\Psi$ means that if we find a solution of \eqref{normsch} with damping rate $\gamma(\eta)$ we have a solution of \eqref{modeeq} with damping rate $\Gamma$ given by,
\begin{equation}
\label{growth}
\Gamma = \frac{r_0^2}{m^2} \frac{D_1}{2 \omega_0} + \frac{1}{2\omega_0}\left( \frac{\omega_H^2 v_A q_0'' r_0^2}{2mq_0^2 R_0} \right)^{2/3} \gamma(\eta)
\end{equation}
\section{Analysis of \eqref{normsch}}
\label{analysis}
\newcommand{\tps}{\tilde{\Psi}}
Having reduced the problem to the simple form of \eqref{normsch} we now proceed to analyse this equation to find $\gamma(\eta)$. Our goal is to find the least damped solution of \eqref{normsch} 
in the absence of a drive. This corresponds to the most unstable/least-damped solution for $\Phi_m$ through \eqref{growth} in the presence of drive.  Examining the form of \eqref{normsch} we 
see that for large $|z|$ the solutions are asymptotically WKB solutions of the form,
\begin{equation}
\tilde{\Psi} \sim \frac{1}{z}e^{\pm i z^3 / 3},
\end{equation}
which are radially ingoing / outgoing waves.  By analogy with the Schr\"odinger equation the $z$-flux of $|\tilde{\Psi}^2|$ is given by $i\left(\tps \partial_z \tps^* - \tps^*\partial_z\tps\right)$, which shows that the outward flux of $|\tps^2|$ at large $z$ is proportional to $z^2 |\tps^2|$. We therefore expect perturbations near the origin to last longest before radiating away to large $|z|$.
It is also seen immediatly from the WKB solutions that as the energy is radiated away from the origin, it quickly reaches ever shorter scales and thus formally our assumption in \eqref{ddrsmall} becomes invalid. 
However, if we assume that there is some (presumably kinetic) dissipation at small scales then we can envisage a situation in which the region around $z=0$ is well described by our model, and it radiates energy away at a rate that is independent of the exact small scale dissipation mechanism. 

In order to calculate the characteristic lifetime of a perturbation we use an initial value code to solve \eqref{normsch}. We discretise both in space and in time, using a standard Crank-Nicolson scheme. In order to mimic a generic small scale damping mechanism at large $z$ at each timestep we multiply the solution by a smooth windowing function that artificially damps fluctuations far from the origin. This provides an absorbing boundary away from the origin allowing energy to radiate away from the origin realistically.

As this is a Schr\"odinger equation with a potential hill, we expect there to be a complete set of solutions with outgoing wave boundary conditions and quantized damping rate. Thus any initial perturbation can be expanded in these functions, and after a few damping times only the least-damped of these solutions (quasimodes) remains. The resulting damping rate is thus independent of initial conditions. We find this damping rate by taking the absolute value of $\tps$ at the origin and fitting $\ln(|\tps|)$ to a linear function of time, after removing the initial transient phase where many modes are still present.

\begin{figure}[htp]
\centering
\includegraphics[width=0.8\textwidth]{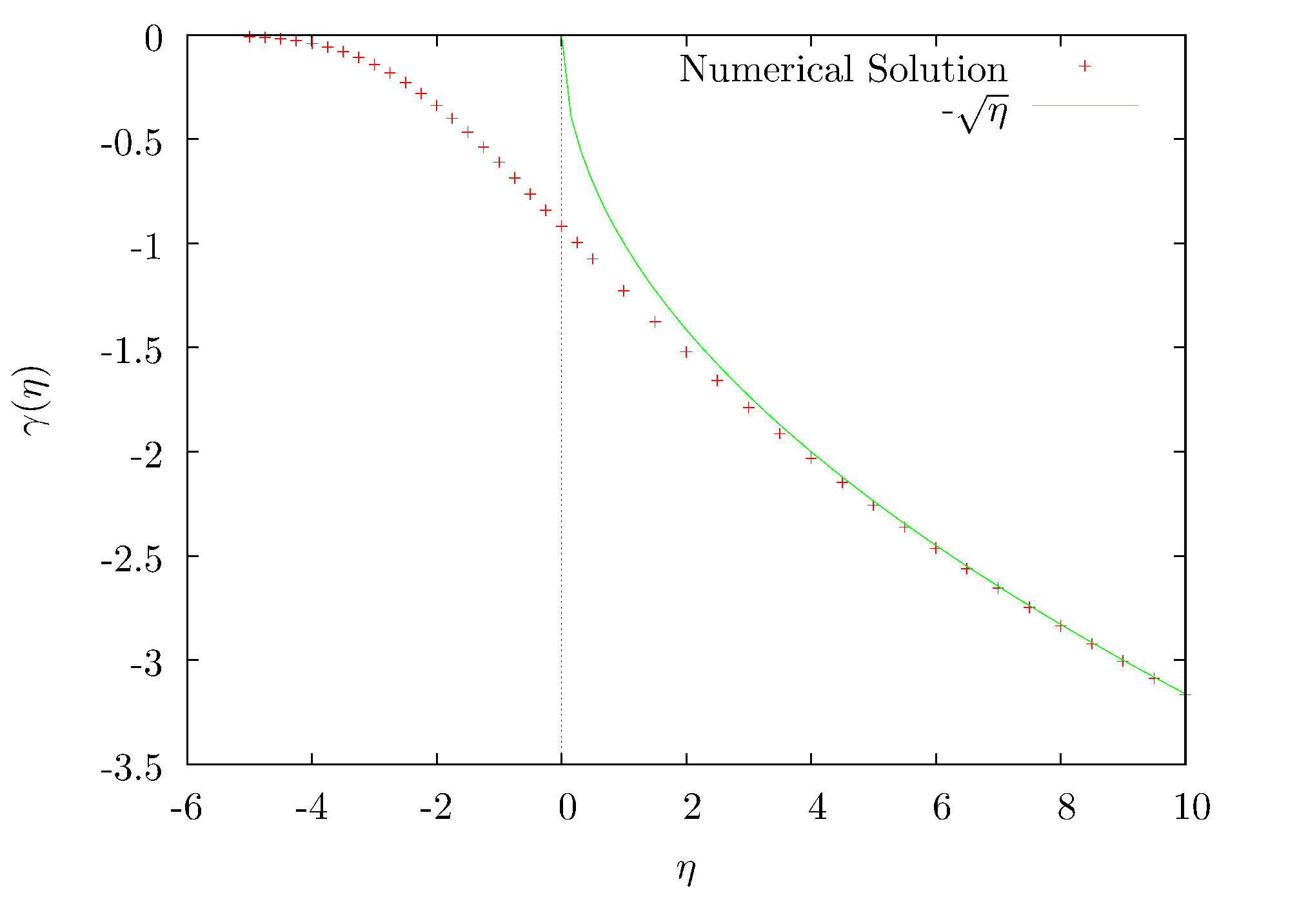}
\caption{Numerical solution for $\gamma(\eta)$ compared to analytic result for large $\eta$ (color online)}\label{fig:numerical}
\end{figure}

We also seek an analytic approximation to this damping rate in order to validate our code. For large positive $\eta$ the problem can be transformed into the standard anharmonic oscillator with a complex quartic term.This problem has been studied before, and so we can use standard results~\cite{bender1969ao} to obtain for large $\eta$
\begin{align}
\label{gammaetalarge}
\gamma(\eta) = -\sqrt\eta \left( 1 - \frac{3}{4} i \eta^{-3/2} + \frac{21}{16} \eta^{-3} \right) .
\end{align}
However, as we only require the real part of $\gamma$ we can neglect the $\eta^{-3/2}$ term. 
We note that the analysis in \tcite{borisvarenna} of \eqref{normsch} for large $\eta$ contains misprints and should read (in our notation) that $\Psi \sim e^{i\sqrt{\eta} z^2 / 2}$ with $\lambda \sim i\sqrt{\eta}$, in agreement with \eqref{gammaetalarge}. As a further check we confirm that in the case of $\eta=0$ we reproduce the ground state of the anharmonic oscillator correctly~\cite{bender1977naw}.
In Fig.~\ref{fig:numerical} we present the analytic approximation for $\gamma$ with the numerical results obtained by our finite difference solver.
We note that the exponential tail for negative $\eta$ gives the small radiative damping of the usual upward sweeping eigenmode.

It is important to compare our results to alternative interpretations of downward sweeping ACs. Whilst we consider a weakly-radiatively-damped mode atop a potential hill, it is possible to construct solutions where kinetic effects provide a potential well for standing waves~\cite{konovalov:4531,gorelenkov:110701} via coupling to kinetic Alfv\'en waves. 

We work in the same regime as \tcite{konovalov:4531} where $\Gamma$, given as the sum of the drive and the continuum damping by \eqref{growth} is not vanishingly small. In this regime, the discrete kinetic modes have a spectrum consisting of multiple eigenmodes with a frequency spacing $\delta \omega$ given by:~\cite{konovalov:4531} 
\begin{equation}
\frac{\delta \omega}{\omega_A} = 2 \rho_i \sqrt{\frac{T_e}{T_i} + \frac{3}{4}} \sqrt{\frac{q''}{q}} \left( \frac{nq}{m} - 1 \right)^{-1/2}.
\end{equation}
We compare this to the damping rate $\gamma_d$ of our longest lived mode in the limit $\eta \gg 1$, i.e. away from the lowest frequency of the cascade or for very small $q''$,
\begin{equation}
\frac{\gamma_d}{\omega_A} = \frac{\omega_H}{\omega_A} \frac{1}{\sqrt{8}} \sqrt\frac{q''}{q} \frac{r}{m} \left(\frac{nq}{m} - 1 \right) ^{-1/2}.
\end{equation}
Thus, if the condition
\begin{equation}
\label{discretization}
 \frac{r}{\rho_i m} \frac{\omega_H}{\omega_A} > 1
\end{equation}
is satisfied, we expect the resulting behaviour to be well described by our continuum approximation. This is because the radiative damping is in reality due to phase mixing that arises from the many different kinetic modes that make up the initial perturbation. The discrete nature of these modes is only apparent on timescales longer than $\delta \omega^{-1}$ by which time the mode has either decayed to zero or grown beyond the applicability of linear theory.
In the opposite limit, where \eqref{discretization} is violated, the results of \tcite{konovalov:4531} have been extended in \tcite{gorelenkov:110701} to find a discrete kinetic mode with a broad radial scale. 

We now consider the condition given by \eqref{discretization} for typical experimental parameters. Using \eqref{omHdef} and neglecting the pressure gradient allows us to estimate $\omega_H$ by
\begin{equation}
\frac{\omega_H}{\omega_A} \approx \sqrt{\frac{1}{m} \frac{\Omega}{\omega_A} \frac{n_h}{n_i} \frac{r}{L_h}},
\end{equation}
where $L_h$ is the density gradient scale length of the fast particles. For typical JET parameters ($\frac{\Omega}{\omega_A} \approx 10^3$, $m \approx 10$, $\frac{n_h}{n_i} \approx 10^{-3}$, $\frac{r_0}{\rho} \approx 100$ and $\frac{r}{L_h} \approx 3$) we find
\begin{equation}
\frac{\omega_H}{\omega_A} \approx \frac{1}{2}
\end{equation}
and 
\begin{equation}
\frac{r}{\rho_i m} \frac{\omega_H}{\omega_A} \approx 5 \gg 1,
\end{equation}
so that the continuum-like description for the mode is suitable for the problem under consideration. Near marginal stability, i.e. for the case $\gamma_L - \gamma_d \ll \gamma_d$, where $\gamma_L$ is the linear drive due to resonant particles, an extra factor $\frac{\gamma_L - \gamma_d }{ \gamma_d} $  appears in Eq.(94) , which may violate the applicability condition for the quasimode approach and a more delicate description of the discrete spectrum would be required.



\section{Comparison With Experiment}
\label{exp}
In relating our results to experimental data, we need to highlight the fact that the {\it downward} sweeping ACs are very rare compared to {\it upward} sweeping ACs.
Since 1997, when magnetic spectrograms became available for every JET discharge, downward sweeping ACs have been detected in only 26 discharges as opposed to more than 5000 discharges where upward sweeping ACs have been observed  ( magnetic probes, O-Mode inteferometry~\cite{sharapov2004mac} and X-Mode reflectometry~\cite{hacquin2007lxr} contribute to this data). 
Due to the scarcity of downward sweeping ACs, some essential plasma diagnostics were unavailable at the time of their observation. In particular, MSE measurements~\cite{brix:10F325} of the q(r)-profile were only made in 5 discharges out of the 26. Although 26 examples of downward sweeping ACs provide very limited statistics they are still indicative of why downward sweeping ACs are so rare.

First, there is no obvious correlation between downward sweeping ACs and the type of energetic particle population in the discharge. The downward sweeping ACs were observed in discharges with NBI only (3 discharges), combined ICRH and NBI (9 discharges), and ICRH only (14 discharges). In the cases of plasmas with ICRH, high-field side (up to $R_{RF} - R_0 \approx -30cm$ in discharge \#56947, t = 3.8 s), low-field side (up to $R_{RF} - R_0 \approx 35cm$ in discharge \#65550, t = 5.2 s), and on-axis ICRH were used. 
It is therefore unlikely that downward sweeping ACs require an extraordinary energetic particle distribution.

Second, downward sweeping ACs are seen in discharges with magnetic fields ranging from 1.8 T (discharge \#72691) to 3.45 T (discharge \#46863). Although no
measurements of ion temperature exist for most of the downward sweeping ACs, the variation of the magnetic field, together with the difference in the types of heating, is
likely to cause a significant spread in Larmor radii across the data base. In the discharges where ion temperature was measured, ion Larmor radius is found to be
about 3 mm, no different from many discharges with upward sweeping ACs.  We then conclude that the downward sweeping ACs are unlikely to be due
to some anomaly in Larmor radius present in all 26 pulses of the data base.

Third, we investigate whether the downward sweeping ACs are associated with a specific type of plasma equilibrium. With plasma currents
varying from 1.1 MA (discharge \#68822 at t = 1.95 s) to 3.5 MA (discharge \#46863 at t = 4.8 s), downward sweeping ACs are observed with $q_{min} \approx 2$, $q_{min}\approx 3$, and even $q_{min} \approx 4$ (discharge \#68822, t = 1.95 s). 
However, equilibrium reconstructions for the discharges where MSE diagnostic data was available show that such discharges have very flat $q$-profiles (See Table\ \ref{tab:data}) . We note that downward sweeping ACs were never observed in JET discharges with strong shear reversal and current holes~\cite{PhysRevLett.87.115001}.
These discharges are typical for JET advanced scenarios and they usually exhibit very clear upward sweeping ACs. One can than conjecture that downward sweeping ACs require low shear values, which are quite rare in JET discharges with $q_{min} >1$.
This conjecture is consistent with what follows from \eqref{growth} and Fig.\ \ref{fig:numerical}: shallow $q$-profiles (small $q''$) minimize continuum damping 
making it easier for the mode to be destabilized and grow to an observable level.
\begin{table}
\centering
\begin{tabular}{ | c || c | c | c | c | c |}
\hline
Pulse \# & $t_{AC}$ s & $q_{min}$ at $t_{AC}$ & $t_{MSE}$ s & $(r^2/q)q''$ & Maximum $s$\\
\hline
66539 & 5.15 & 2 & 4.96 & 0.14 & -0.06 \\
66550 & 5.2 & 2 & 4.96 & 0.4 & -0.1 \\
72961 & 2.7 & 3 & 2.21 & 0.9 & -0.22 \\
72818 & 3.9 & 2 & 3.06 & 0.4 & -0.11 \\
74896 & 2.25 & 2 & 2.4 & 0.35 & -0.07\\
\hline
\end{tabular}
\caption{Discharges with downward sweeping ACs, for which MSE measurements are available}
\label{tab:data}
\end{table}

Toroidal mode numbers were determined from Mirnov coil data for 19 discharges containing downward sweeping ACs. All these numbers were positive and lie in the range $n=1$ to $n=5$. 
It is noteworthy that upward sweeping ACs also have only positive toroidal mode numbers~\cite{sharapov2001mhds}. The absence of negative toroidal mode numbers agrees with the theoretical 
conclusion that the corresponding modes are stable and damped.

An idealized model of quiescent plasma with a population of hot ICRF heated ions and a shallow ($|s| \ll 1$) $q$-profile 
appears to be relevant to reverse-shear discharges with downward sweeping ACs.
As follows from \eqref{etadef}, these modes must have positive values of $\eta$. The time evolution
of $\eta$ is dominated by that of $nq-m$. We thus notionally hold all other parameters constant and 
examine what happens as $q$ (or equivalently $\eta$) decreases with time. As the continuum damping decreases with $\eta$
the mode eventually becomes unstable.  If the critical value of $\eta$ is above zero we expect to see a downward sweeping trace until $\eta$ reaches zero whereupon the mode becomes the
Geodesic Acoustic Mode\cite{breizman2005ppe}. As $\eta$ decreases further to negative values, we expect to see an upward sweeping trace until the mode reaches the TAE frequency.

\begin{figure}
\centering
\includegraphics[width=80mm]{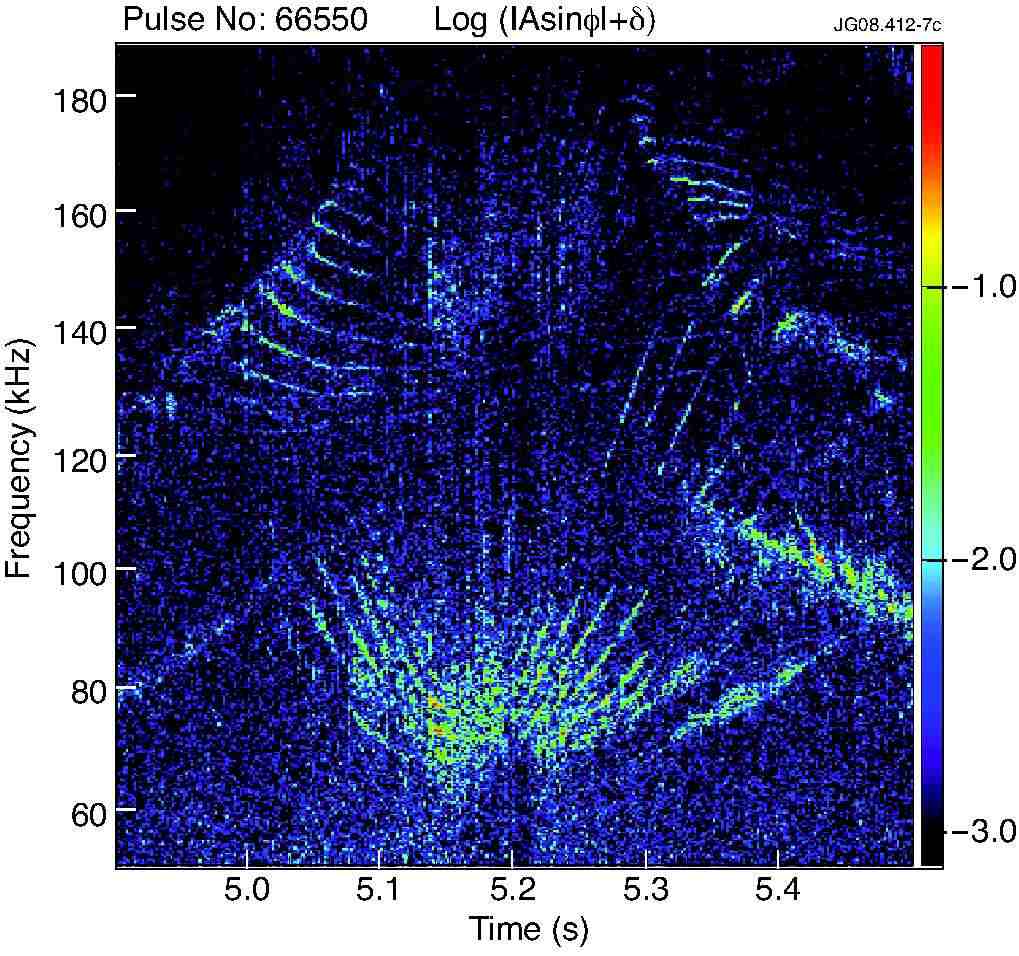}
\caption{Spectrogram of O-Mode interferometry from JET pulse \#66550 showing ACs (color online)}
\label{omode}
\end{figure}

\begin{figure}
\centering
\includegraphics[width=80mm]{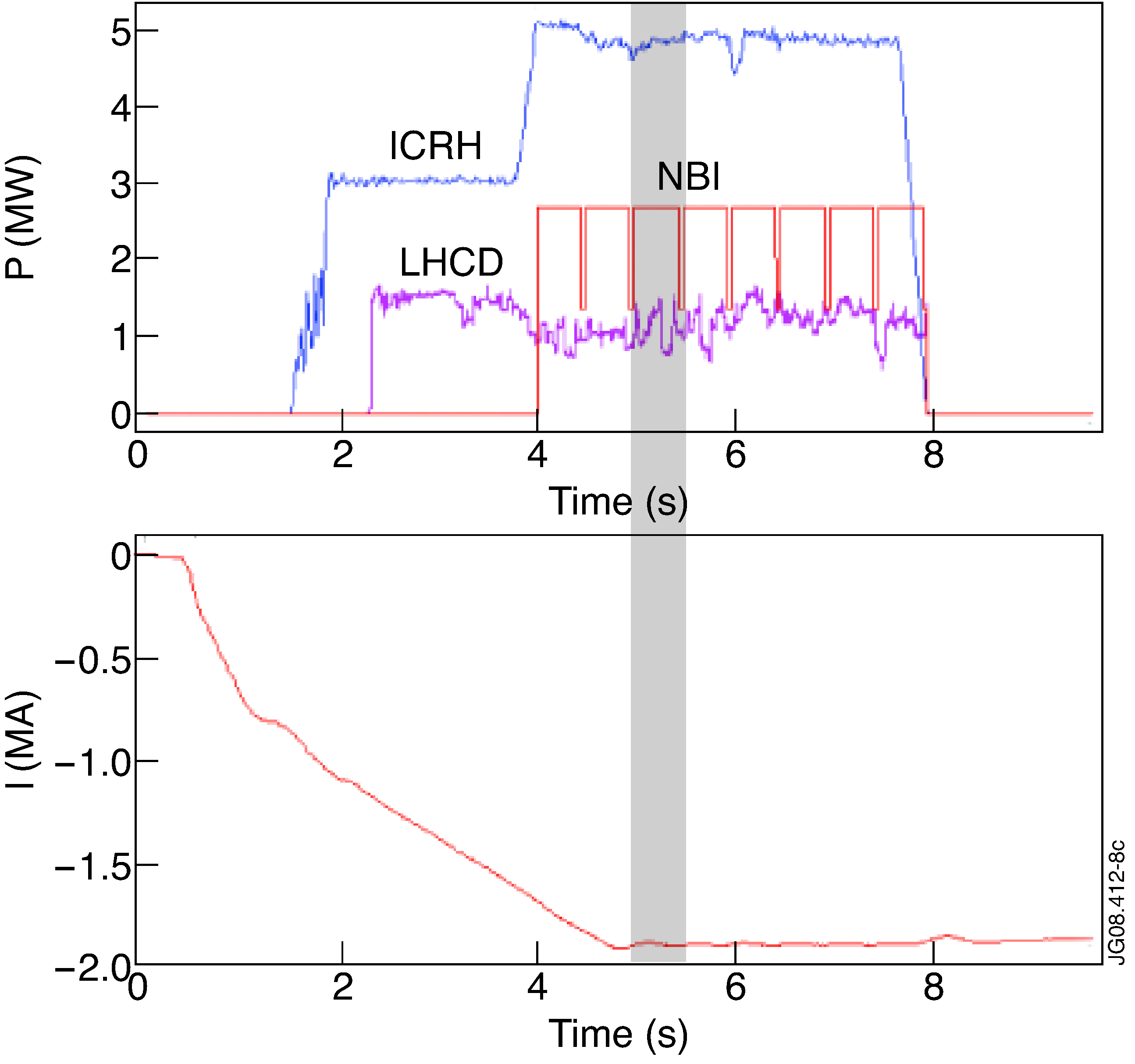}
\caption{ICRH, NBI, LHCD power waveforms, and plasma current signal for JET Pulse \#66550, shaded time interval corresponds to the time shown in Fig.\ \ref{omode} (color online)}
\label{power}
\end{figure}

\begin{figure}
\centering
\includegraphics[height=80mm]{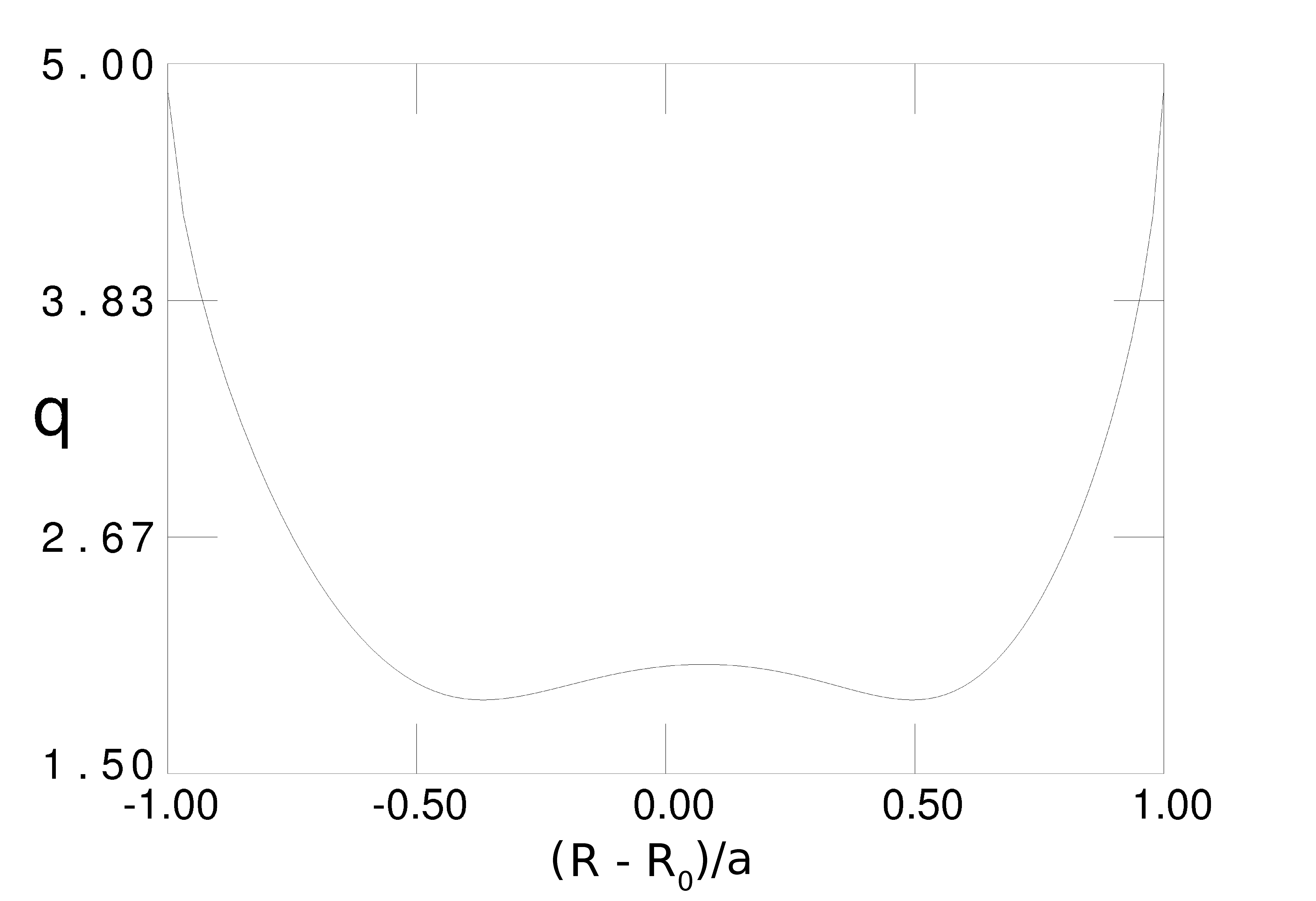}
\caption{Reconstructed $q$ profile, from MSE and polarimetry data for JET pulse \#66550 }
\label{qprof}
\end{figure}

We now consider an example of downward-sweeping ACs shown in Figure~\ref{omode}. Figure~\ref{power} shows the corresponding power waveforms for the ICRH, NBI and Lower-hybrid current drive (LHCD) systems in JET discharge \#66550 with toroidal field $B_T = 2.46\ T$ and plasma current $I_{pla} \lesssim 1.9 MA$. It is typical for scenarios aiming at the creation of an internal transport barrier that the inductive current drive is still ramping up (see Figure~\ref{power}) when the main heating power is applied. This creates the non-monotonic $q$-profile shown in Figure~\ref{qprof} (reconstructed from MSE data), with $q_{min} = 2.1$ at $t=4.96s$ when the MSE data was taken. 

This profile has a broad region of very low shear ($|s| < 0.1$ in the shear-reversed region) with $r^2 q'' / q = 0.4$ at the shear  reversal point.

\begin{figure}
\centering
\includegraphics[height=80mm]{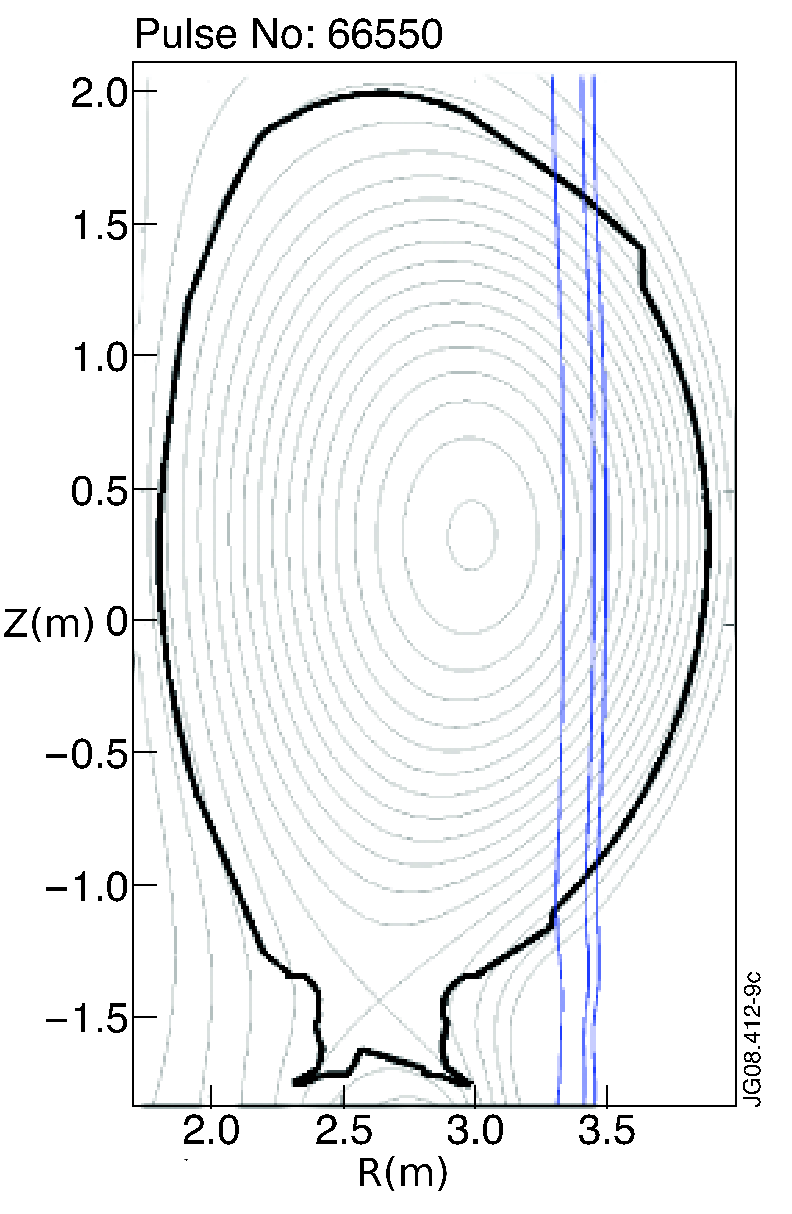}
\caption{ICRH resonant surfaces at time of AC, with reconstructed flux surfaces shown (color online)}
\label{ICRH}
\end{figure}

Figure~\ref{ICRH} shows the resonant surfaces for the ICRH antennas at the time of observation. Most of the RF power was deposited at $R = 3.5m$. We thus expect a hollow profile of fast ions in the region $3m < R < 3.5m$. The \verb#SELFO# code~\cite{hellsten2004efd} confirms this expectation, predicting a hollow profile for the fast particle energy density. As seen from \eqref{eq81}, an energetic particle population with such a hollow profile and $\theta_b < \pi / 2$ provides a drive for AC quasimodes making them observable.

It is important to emphasize that the quasimodes discussed here do not require a potential well, which distinguishes them from the AC eigenmodes discussed in \tcite{breizman2003tae}. The drive calculated here is destabilizing only for positive toroidal mode numbers, in the typical off-axis ICRH scenarios. Downward sweeping ACs have also been observed in on-axis heated discharges. In order to apply our theory to this case we need to include a broader profile in bounce angle in our derivation. The generalization of \eqref{dpfinale} to a localized distribution with finite width is fairly simple, and one can see quickly that it will be a weighted average of \eqref{dpfinale} over the range of possible bounce angles. However as $K_1$ is larger in the high field side region (by one power of $n$) we can neglect any contrbution from the low-field-side region and conclude that, for on or near axis heating, positive $n$ modes are unstable if $\pd{F}{P_\varphi} < 0$ at the zero shear point. This agrees with experiment where, unless the zero shear point is very close to the magnetic axis, we indeed expect the profile to drop off radially ($\pd{F}{P_\pht} < 0$) at the zero shear point. Due to the scarcity of downward sweeping modes on JET, the inability to run with hollow profiles and high-field-side IRCH and the scarcity of toroidal mode number data it is not possible to completely verify the predictions made after \eqref{eq81}. However none of the experimental results on downward sweeping modes from JET contradict our predictions. 

\section{Conclusion}
In this paper we have used the action-angle formalism for the bounce-precessional motion of the fast ICRH-accelerated ions to calculate the resonant response of the ions to shear Alfv\'en perturbations without making the deeply trapped orbit approximation used previously~\cite{fulop1996fow,zonca:4600}. By exploiting the characteristic features of Alfv\'en Eigenmodes, we reduce the complex orbit integrals to local contributions from a few critical points.

Using this local formulation of the energetic ion response we calculate the resonant ion drive for Alfv\'en Cascades (ACs) and AC quasimodes. Based on the expression for the drive, and an asymptotic expression for the radiative damping, we conclude that the AC quasimode should be observed predominantly in weakly reversed shear configurations and that, in such a configuration, it can explain downward sweeping Alfv\'en Cascades below the TAE frequency. Furthermore, we conclude that the quasimode should only be excited with positive mode numbers in experimentally relevant plasma configurations. 

We  reinforce this conclusion via a comparison to experiments carried out on JET. A database of JET pulses where quasimodes were observed is compiled and examined, along with advanced equilibrium reconstruction to determine the $q$-profile. The results obtained demonstrate that the observation of downward sweeping Alfv\'en Cascades is indeed linked to the presence of a weakly reversed shear plasma. In the few cases where mode number information is available only positive mode numbers are excited, agreeing qualitatively with our predictions for the resonant excitation mechanism due to trapped energetic ions.

\begin{acknowledgments} 
During this work I.G.A.\ was supported by a CASE EPSRC studentship in association with UKAEA Culham. The authors are grateful for the travel support provided to  B.B. by the Leverhulme Trust (UK) International Network for Magnetized Plasma Turbulence.
This work was funded in part by the UK Engineering and Physical Sciences Research Council, by the European Communities under the contract of Association between EURATOM and UKAEA, and by U.S. Department of Energy Grant no. DE-FG03-96ER-54326.

We thank T. Johnson for modelling the ICRH-accelerated particle populations. We would also like to thank M. Brix and the JET MSE diagnostic team for providing $q$-profile measurements and reconstructions.
This work was carried out within the framework of the European Fusion Development Agreement, the  views and opinions expressed herein do not necessarily reflect those of the European Commission. 

\end{acknowledgments}

\appendix
	\section{Evaluation of \eqsref{kdef}{kextdef}}
\label{apintegrals}
In this Appendix we evaluate the integrals $K(l,m)$ and $K^{ext}(l,m)$ defined in \eqsref{kdef}{kextdef} respectively in the high-$n$ limit.
These integrals relate harmonics in straight-field-line variables in which the fields are naturally expressed to the action-angle variables $\phb$ and $\thb$ adapted to the particle orbits.

The major simplifying assumption that allows us to proceed in this appendix is the high mode number limit $n,m \rightarrow \infty$, however 
we immediatly observe that $k_\parallel q R \sim 1$ and so $nq -m$ doesn't scale with $n$ as $n \rightarrow \infty$.
We also neglect the small difference $\tht - \thh$ unless it is multiplied by the large mode number $n$ or $m$.

We note the symmetry in the definitions of $K$ and $K^{ext}$ and so manipulate $K^{ext}$ using \eqsref{phbdef}{thbdef} to find
\begin{align}
K^{ext}(l,m) = \sum_\sigma\frac{1}{2\pi}\int\limits_{-\theta_b}^{\theta_b} d\theta \left|\frac{d \thb}{d \theta}\right| e^{-i n G(\theta) \chi_1(\bar r) } e^{-i(nq-m)\hat{\theta}} e^{i \frac{n}{\omega_{\thb}} \left(\omega_{\phb} - \chi_2(\bar{r}) \right) \thb } e^{-il\thb}.
\end{align}
Now we can convert this to an integral over $\thb$,
\begin{align}
K^{ext} = \frac{1}{2\pi}\int\limits_{-\pi}^{\pi} d\thb e^{-i n G(\theta) \chi_1(\bar r) } e^{-i(nq-m)\hat\theta} e^{i \frac{n}{\omega_{\thb}} \left(\omega_{\phb} - \chi_2(\bar{r}) \right) \thb } e^{-il\thb},
\end{align} 
and comparing this to the definition of $K$ in \eqref{kdef} we find
\begin{align}
K^{ext}(l,m) = K^*(l,m),
\end{align}
with $^*$ denoting the complex conjugate.

Proceeding now with the explicit evaluation of $K$ we use \eqref{kdef} and \eqsref{phbdef}{thbdef} to obtain
\begin{align}
K(l,m) &= \frac{1}{4\pi^2} \int\limits_0^{2\pi} d\phb \int\limits_{-\pi}^{\pi} d\thb e^{i(n\phh - m \thh)} e^{-in\phb + i l\thb},\\
		  \label{A4}
          &= \frac{1}{2\pi} \int\limits_{-\pi}^{\pi} e^{i {\hat{l}} \thb}  e^{i n G(\theta) \chi_1(\bar r) + i n\chi_2 \thb} e^{i(nq-m)\tht} d\thb,
\end{align}
where $\hat{l} \equiv l - n \frac{\omega_\phb}{\omega_\thb}$. If we are near resonance then $\hat{l}$ does not scale with $n$.
This integral is now of the form:
\begin{align*}
\int\limits_{x_1}^{x_2} g(x) e^{i n f(x)} dx \mbox{ as $n \rightarrow\infty$},
\end{align*}
for which a standard expression exists,\cite{bender1999amm}
\begin{align}
\label{asympt}
\int\limits_{x_1}^{x_2} g(x) e^{i n f(x) } \mathrm{d} x \sim -\frac{ig(x_2) e^{inf(x_2)}}{n f'(x_2)} + \frac{ig(x_1) e^{inf(x_1)}}{n f'(x_1)} + \sum\limits_{c : f'(c)=0} g(c) e^{inf(c) + i \pi \mu(c)/4} \sqrt\frac{2\pi}{|n f''(c)|}, 
\end{align}
where $\mu(c) \equiv n f''(c)/| n f''(c)|$. From this expression we note that such integrals are dominated by contributions from stationary phase points where $f'(x) = 0$ and from the ends of the interval if no such points exist. We will consider both contributions as stationary phase points do not always exist. The result should be interpreted as providing the leading term. If a stationary phase point exists then the endpoint contributions must be ignored.

To find the stationary phase points, which we denote by $\theta = \theta^*$, we solve,
\begin{equation}
\pd{}{\thb} \left( G(\theta) \chi_1(\bar{r}) + \chi_2 \thb \right) = 0,
\end{equation}
which we can rearrange to give
\begin{align}
\label{t1}
\pd{G}{\tht} + \frac{\chi_2}{\chi_1} \pd{\thb}{\tht} = 0,
\end{align}
where $\pd{\thb}{\tht}$ is related to $\pd{G}{\tht}$ by \eqsdash{dthetadtheta}{vpar} , and the ratio $\chi_2/\chi_1$, in the low-shear limit, becomes
\begin{equation}
\frac{\chi_2}{\chi_1} = \frac{\cos \tht_b}{R_0\omega_\thb q} \sqrt{\frac{2P_\zeta\Omega}{M}\frac{\bar{r}}{R_0} }.
\end{equation}
\eqref{t1} can then be rewritten as
\begin{equation}
2\left(\pd{G}{\tht}\right)^2 = \cos \tht_b,
\end{equation}
which using \eqref{badger} finally gives
\begin{equation}
\cos \tht^* = \frac{1}{2} \cos \tht_b.
\end{equation}
If, without loss of generality, we let $\tht^*$ be the positive solution of this equation we have stationary phase points at $\tht = \pm\tht^*$, and thus 
four stationary phase points in total as each $\tht$ occurs once on each branch of $\thb(\tht)$ (see Eqs.\ (\ref{thbdef}),(\ref{Gb1}) and (\ref{Gb2})).

Dividing the $\thb$ integral in \eqref{A4} into two intervals, corresponding to the two branches of $\thb(\tht)$ (see \eqsref{dthetadtheta}{dthetadtheta2}), we have 
\begin{align}
K(l,m)  &= \frac{1}{2\pi} \left( I + J \right) \\
		I &= \int\limits_{-\pi}^0 e^{i(nq-m){\tht}} e^{i\hat{l}\thb} e^{in G(\theta) \chi_1(\bar{r}) +i n \chi_2 \thb } d\thb \\
		J &= \int\limits_0^{\pi} e^{i(nq-m){\tht}} e^{i\hat{l}\thb} e^{in G(\theta) \chi_1(\bar{r}) +i n \chi_2 \thb} d\thb
\end{align}
We now calculate $I$ and $J$ in the limit of $n \rightarrow \infty$.

Applying \eqref{asympt} to $I$ we find the following contribution from the endpoints $\theta=-\theta_b$,$\thb=-\pi$
and $\theta= \theta_b$,$\thb = 0$:
\begin{align}
e^{-i(nq -m)\theta_b} e^{-i\hat{l}\pi- i n\chi_2 \pi -i\chi_1 G(-\theta_b)} \frac{i}{ n \chi_2} - e^{i(nq -m)\theta_b} \frac{i}{ n \chi_2}
\end{align}
At the stationary phase point, $\theta^*$, we have
\begin{align}
\chi_1 \left.\frac{d^2 G}{d \thb^2}\right|_{\theta=\theta^*} = \frac{2 P_\zeta}{M} \frac{\Omega \bar r \chi_1}{q^2 R_0^3 \omega_\thb^2} \left( - \sin \tht^* \sqrt{\cos \tht^* - \cos\tht_b}\right) \equiv G^*.
\end{align}
Noting that $G^*$ is odd in $\theta^*$, we have the contribution to $I$ from $\theta = +\theta^*$ and $\theta = -\theta^*$,
\begin{equation}
\begin{split}
e^{i(nq-m)\tht^*}& e^{in G(\theta^*) \chi_1 -i n (\hat{l}+\chi_2) (\thb^*) + i \pi\mu / 4} \sqrt\frac{2\pi}{|n G^*|} \\
 &+ e^{-i(nq-m)\tht^*} e^{in G(-\theta^*) \chi_1 -i n (\hat{l}+\chi_2)(\pi-\thb^*) - i \pi\mu / 4} \sqrt\frac{2\pi}{|n G^*|}.
 \end{split}
\end{equation}
Where we introduce $\thb^*$ which is the absolute value of $\thb$ when $\tht = \tht^*$, also $\mu = n G^* / |nG^*|$. Thus the two stationary phase points in the interval $\thb \in [-\pi,0]$ are
$\thb = \thb^* - \pi$ and $\thb = -\thb^*$. Combining the above results gives,
\begin{align}
I = &  e^{-i(nq -m)\theta_b} e^{-i\hat{l}\pi- in\chi_2 \pi -in\chi_1 G(-\theta_b)} \frac{i}{n \chi_2} -  e^{i(nq -m)\theta_b} \frac{i}{n \chi_2}\\\nonumber
&+e^{i(nq-m)\theta^*} e^{-i(n\chi_2 + \hat{l})\left( \thb^*\right)} e^{in G(\theta^*) \chi_1 + i \pi\mu / 4} \sqrt\frac{2\pi}{|n G^*|} \\\nonumber
&+ e^{-i(nq-m)\theta^*} e^{-i(n\chi_2 + \hat{l})\left(\pi - \thb^*\right)} e^{in G(-\theta^*) \chi_1 - i \pi\mu / 4} \sqrt\frac{2\pi}{|n G^*|}.
\end{align}
In this expresssion and hereafter $G(\theta)$ is understood to be evaluated on the branch where $G(\theta) < 0$.
We perform the same analysis for $J$ to give:
\begin{align}
J = 
& e^{i(nq -m)\theta_b} \frac{i}{n \chi_2}
-  e^{-i(nq -m)\theta_b} e^{i\hat{l}\pi+ in\chi_2 \pi -in\chi_1 G(-\theta_b)} \frac{i}{n \chi_2} \\ \nonumber
&+e^{i(nq-m)\theta^*} e^{i(n\chi_2 + \hat{l})\left(\thb^*\right)} e^{-in G(\theta^*) \chi_1 - i \pi\mu / 4} \sqrt\frac{2\pi}{|n G^*|} \\\nonumber
&+ e^{-i(nq-m)\theta^*} e^{i(n\chi_2 + \hat{l})\left(\pi -  \thb^*\right)} e^{-in G(-\theta^*) \chi_1 + i \pi\mu / 4} \sqrt\frac{2\pi}{|n G^*|}.
\end{align}
We can then calculate $K(l,m)$
\begin{equation}
\begin{split}
K(l,m) &= \frac{1}{2\pi} \left(I+J\right) \\
  &= \frac{1}{2\pi} \frac{2}{n\chi_2} e^{-i(nq-m)\theta_b-in \chi_1 G(-\theta_b)} \sin\left(\hat{l}\pi +n\chi_2\pi\right)  \\
  + 2\frac{1}{2\pi}\sqrt\frac{2\pi}{|n G^*|} 
&  \left\{
e^{i(nq-m)\theta^*} \cos\left({(n\chi_2 + \hat{l})\thb^* - n G(\theta^*) \chi_1 - \pi\mu / 4}\right)\right.\\ 
&\left.
+e^{-i(nq-m)\theta^*} \cos\left((n\chi_2 + \hat{l})\left(\pi - \thb^*\right) - n G(-\theta^*) \chi_1 + \pi\mu / 4\right)
  \right\}
\end{split}
\end{equation}

To calculate products of $K$ and $K^*$ in a simple form we employ an averaging over $\bar{r}$, which is permitted by \eqref{ddrsmall} as $\Phi$ doesn't have small scale structure in $\bar{r}$.
Thus we drop all terms containing expressions of the form $e^{inf(\bar{r})}$ leaving only those terms which vary slowly in the radial direction.
This procedure gives,
\begin{align}
\label{llama}
K(l,m) K^*(l,m+s) =  \frac{e^{-is\theta_b} }{2 \pi^2 n^2 \chi_2^2} + \frac{2}{|n\pi G^*|} \left( e^{is\theta^*} + e^{-is\theta^*} \right),
\end{align}
from which we can calculate $K_1$ and $K_2$ as defined in
\begin{align}
\label{K1def}
K_1 = \sum\limits_{s = \pm 1} K(l,m)K^{ext}(l,m+s)   = \frac{\cos\theta_b }{\pi^2 n^2 \chi_2^2} + \frac{8\cos\theta^*}{\pi |n G^*|}, \\
\label{K2def}
K_2 = \sum\limits_{s = \pm 1} s K(l,m)K^{ext}(l,m+s) = - i \frac{\sin\theta_b}{\pi^2 n^2 \chi_2^2} .
\end{align}

\end{document}